\documentclass[ final ] {aipproc}

\usepackage{times}
\usepackage{graphicx}

\newcommand{\msolar}{M_\odot}
\newcommand{\lsolar}{L_\odot}
\newcommand{\mum}{$\mu$m }
\newcommand{\kms}{\,\mbox{km s$^{-1}$}}

\newcommand{\OII}{[{O}{\small II}]}
\newcommand{\OIII}{[{O}{\small III}]}
\newcommand{\Ha}{H$\alpha\,$}

\newcommand{\HI}{{H}{\small I}}
\newcommand{\HII}{{H}{\small II}}

\newcommand{\eg}{\emph{e.g.}}

\layoutstyle{6x9}

\begin{document}

\title{Modelling the UV to sub-mm SED of \\ Starburst Galaxies}

\author{Michael A. Dopita}{
  address={Research School of Astronomy \& Astrophysics,
  The Australian National University, \\ Cotter Road, Weston Creek, ACT 2611, Australia}
}

\begin{abstract}
 This paper reviews the theoretical modelling of starburst galaxies with particular emphasis on the importance of the gas-phase physics in determining the form of the continuum SED and the  emission line spectrum. Both line and continuum diagnostics have been used as star formation indicators. Here, we examine from the point of view of the theory how reliable each of these indicators may be, and to what physics of the galaxy they are sensitive, with the aim of decreasing the errors associated with measuring the star formation history of the universe.
 \end{abstract}

\maketitle


\section{1. Introduction}

What motivates our theoretical attempts to model the SEDs of galaxies? Put another way, what do we hope to learn from such an exercise? Firstly, we should recall that the bolometric luminosity of a starburst galaxy is dominated by the young stars it contains. Thus regardless of how much or how little of this luminosity is reprocessed through the dusty interstellar medium either through thermal emission in the IR of dust grains, through fluorescent processes or through heating and re-emission in an ionized medium, the pan-spectral SED tells us what the star formation rate currently is, or has been in the recent past. The first objective of pan-spectral SED modelling is therefore to be able to reliably infer star formation rates in galaxies and to provide likely error estimates using observational data sets which may in practice be restricted to only certain emission lines or spectral features. 

The importance of being able to reliably measure star formation cannot be under-estimated, since the star formation history of the universe is a fundamental indicator of galaxy evolution. The famous Lilly-Madau Plot \cite{Madau96} provides this quantity on a global scale. This plot shows the star formation rate (in solar masses per year) per (cosmology corrected) co-moving volume (Mpc$^3$) vs. the redshift, $z$. As shown notably by Ascasibar et~al.\cite{ascasibar02}, an unacceptable scatter attaches not only to the observational estimates, and also to theoretical estimates based on the $\Lambda$CDM cosmology. For the observational data, much of this scatter results from uncertain corrections for the absorption by dust, and from the application of different methods with different observational biases and uncertainties. For the theoretical curves the uncertainties are mostly due to cosmic variance, the choice of the ``prescription'' of  star formation which has been adopted, and to the treatment of feedback between the star formation and the interstellar medium.

Beyond the global picture of the star formation history of the universe provided by the Madau Plot, there is increasing interest in deriving the star formation rates of individual galaxies at high redshift to determine how galaxies were assembled as a function of environment and of time. At high redshift, massive dusty protogalaxies have been  observed to be forming stars at many thousands of solar masses per year \cite{reuland03}.  If continued, such bursts would convert all the galactic gas into stars in roughly a dynamical timescale; $\sim10^8$ years. By contrast, there is increasing evidence that the Lyman break galaxies represent a more moderate (and probably much less massive) class of object with, in many cases, little dust obscuration and star formation rates $\sim 50 \msolar$yr$^{-1}$. In order to interpret the SEDs of such distant objects, we first need to understand how the form of the SED is controlled by the interstellar physics and the geometry of the stars with respect to the gas. To do this, we would be best advised to first study and model nearby star-bursting objects such as Arp~220 in which we can better resolve the structure of a starburst region, and for which we frequently have much better quality data over more wavebands than we have for most high-redshift objects. Once understood, such objects can then be used to gain physical insight into these more distant sources

Such theoretical studies of nearby objects provide the second motivation for SED modelling; to gain insights into the physical parameters of starburst galaxies, such as the stellar populations, the atomic and molecular gas content, the star and gas-phase metallicities, the physical parameters such as pressure or mean density of their interstellar media, and the nature of the interstellar dust, its composition and spatial distribution. These physical parameters then drive insight into the physical processes which control them. These include as supernova processing, shock-induced star formation, dust grain processing, synchrotron losses as well as many others.

\section{2. Star Formation Indicators}

Almost any feature in the SED of a starburst galaxy could, in principle, be used as a star formation rate (SFR) indicator, provided the appropriate bolometric correction can be worked out. In many cases, however, this is by no means a trivial exercise. Here we briefly discuss some of the more commonly-used techniques to determine the SFR in galaxies. Others exist (for example, the X-ray flux), but these are not yet sufficiently well understood from a theoretical point of view to warrant inclusion here.

\paragraph{\bf The optical / UV continuum} For those techniques which depend upon optical or UV data \eg \cite{Calzetti01}, the effects of dust obscuration in and around the star-forming regions are particularly severe, and make it mandatory that we understand the dust attenuation physics, which is discussed extensively by Fischera here \cite{Fischera04b}. In principle, provided that the initial mass function (IMF) is invariant, the intrinsic UV luminosity should scale directly as the star formation rate. From stellar spectral synthesis models, the star formation rate is given in terms of the 1500\AA\ flux by \citep{Kennicutt98,Panuzzo03}:

\begin{equation}
\left[SFR_{{\rm UV }} / \rm {M_{\odot}~yr^{-1}}\right]
=(1.2 -1.4)\times 10^{-28}\left[ L_{{\rm UV }} / {\rm erg~s}^{-1}~{\rm Hz}^{-1}\right] .  \label{UV}
\end{equation}

\paragraph{\bf The far-IR continuum} The use of the far-IR emission from warm dust associated with the star formation region may provide a more reliable estimate of the SFR \cite{Dwek98, Gispert00}. However, star formation rates derived in this way most frequently assume that the dust is effectively acting as a re-processing bolometer wrapped around the star forming region, and that cool and old stars do not provide too much of a contribution.

In any real galaxy, old stars exist, and the stars are neither unobscured by dust, and nor are they totally obscured \cite{Popescu00}. This degree of obscuration  is one of the many ``geometrical'' factors which render the derivation of the absolute SFR quite uncertain. For example, \citet{Inoue02} includes a number of these geometrical factors in the following formula for the IR luminosity of a galaxy:

\begin{equation}
L_{{\rm IR}} = L_{{\rm Ly\alpha}} + (1-f)L_{{\rm LC}} +
\epsilon L_{{\rm UV}} + \eta L_{{\rm old}} \label{Inoue}
\end{equation}

This allows a fraction, $(1-f)$, of the ionizing flux to be absorbed by dust in the \HII\ region, assuming that all of the Lyman-$\alpha $ photons are multiply scattered by the gas and ultimately absorbed by dust in the surrounding \HI\ region, that a fraction, $\epsilon $, of the non-ionizing UV photons are absorbed, and that a fraction $\eta $ of the radiation field produced by the old stars is also absorbed by dust. Photoionization models of \HII\ regions show that, typically, the Lyman-$\alpha $ flux is of order 30\% of the total stellar flux absorbed by the nebula, so that globally, the dust is fairly efficient in capturing and re-radiating the flux originally radiated by the star in the Lyman continuum.

The dust obscuration is observed to increase both with galaxian mass \cite{Kewley02}, and with the absolute rate star formation \cite{Buat99,Adelberger00,Dopita02,Vijh03}. This can be understood as the combination of the degree of chemical evolution of the interstellar medium, which leads to a low dust to gas ratio in dwarf galaxies, and of the operation of the Schmidt Law \cite{Schmidt59} of star formation, which couples the star formation rate per unit area to the gas surface density, and therefore, for a given dust to gas ratio, to the dust obscuration, provided that the surface rate of star formation is coupled to the global rate of star formation, which appears to be most often the case. Recent data \cite{Kennicutt98, Misiriotis04} gives $\Sigma_{\rm SFR} \propto \Sigma_{\rm gas}^n$, with $n \sim 1.3-1.6$ .

\paragraph{\bf The radio continuum} The radio continuum has also been used as a star formation rate indicator. Observationally, there is an extraordinarily close correlation between the 60$\mu$m infrared continuum and the radio 1.4~GHz continuum of star forming galaxies. This linear correlation spans  $\sim 5$  decades of magnitude with less than 0.3~dex dispersion \cite{Yun01, Wunderlich87}. The mean relationship between the 60$\mu$m flux and the 1.4~GHz continuum is:

\begin{equation}
\left[{L_{\rm 1.4GHz}} / {\rm W~Hz^{-1}}\right]=10^{12}\left[{L_{\rm 60\mu m}} / {L_\odot}\right]
\label{radIR}
\end{equation}

In star forming galaxies at 1.4~GHz, the non-thermal emission by relativistic electrons dominates by at least an order of magnitude over the free-free emission \citep{Condon92}. Therefore, and somewhat remarkably, this relationship couples a purely thermal process with a non-thermal process, over many decades of flux, and locally, within individual galaxies. 

If the lifetime of the synchrotron electrons is short compared with the evolution timescale of the starburst, then as \citet{Bressan02} have shown, the synchrotron emissivity acts as a bolometer of the supernova rate. This works because the relativistic luminosity is ultimately derived from the Fermi acceleration process in supernova shocks, which for constant efficiency should be proportional to the star formation rate. Specifically, the non-thermal luminosity is given by:

\begin{equation}
L_{\rm NT}(\nu) \propto \dot{M}_{*} B^{\alpha-1}\nu^{-\alpha} \label{NT1}
\end{equation}
where $\dot{M}_*$ is the total star formation rate. Since the observed spectral index lies in the range, $0.5 > \alpha > 1.0$, there is only a very weak dependence on the magnetic field in this case. To the extent that the FIR luminosity is providing a bolometric indication of the SFR, this seems to provide the most natural explanation of the FIR :  radio correlation in starburst galaxies. This will be especially true when the pressure in the interstellar medium is high, since provided that the magnetic pressure is in equipartition with the gas pressure, the synchrotron losses are strongly encouraged by the high interstellar magnetic field.

\paragraph{\bf Recombination emission lines} A fairly direct and extensively-used technique is to measure hydrogen recombination line fluxes. Provided that the \HII\ region can absorb all the EUV photons produced by the central stars, this should be a reliable technique, since in this case the flux in any hydrogen line is simply proportional to the number of photons produced by the hot stars, which is in turn proportional to the birthrate of massive stars. This relationship has  been well calibrated at solar metallicity for the H$\alpha $ line. In units  of M$_{\odot}$~yr$^{-1}$, the estimated star formation rate is given by \citep{Dopita94,Kennicutt98,Panuzzo03}:

\begin{equation}
\left[SFR_{{\rm H\alpha }} / \rm {M_{\odot}~yr^{-1}}\right]
=(7.0-7.9)\times 10^{-42}\left[ L_{{\rm H\alpha }} / {\rm erg.s}^{-1}\right] .  \label{2}
\end{equation}

Provided that the star-formation regions are resolved, their Balmer decrements can be used to estimate the absorption in any foreground dust screen.  However, it is possible that some star forming regions are  completely obscured, even at H$\alpha$. This problem can be avoided by observing at infrared wavelengths,  in Br$\alpha ,$ for example. A further complication is that the dust content of the nebula is metallicity-dependent, and therefore an appreciable fraction of the hydrogen ionizing photons may be absorbed by dust in high metallicity \HII\ regions.  This possibility, first seriously quantified by \citet{Petrosian78},  has been discussed by a  number of authors since and its effect has been investigated and quantified (as far as is possible by direct  observation) in a series of recent papers by Inoue and his collaborators \cite{Inoue00, Inoue01a, Inoue01b}.  Dopita \cite{Dopita03} has shown that this effect increases in importance in the more compact  \HII\ regions, and  may lead to errors as  high as $\sim 30$\% in estimates of the global SFR. Indeed, in ultra-compact \HII\ regions, as much as $\sim 90$\% of the ionising photons may be lost to dust in the ionized nebula. However, the magnitude of this effect is a function of the product of the metallicity, $Z$ and the dimensionless ionization parameter, ${\cal U}$, both of which can in principle be determined by emission line diagnostics.

\paragraph{\bf \OII\ emission lines} The Balmer lines shift out of the visible range by $z > 0.4$. The other strong emission lines accessible at higher redshift are the \OII\ and \OIII\  lines. Of these, the \OIII\ lines are known to be very sensitive to the ionization parameter, and therefore the \OII\ $\lambda3727,9$\AA\ lines have been used of necessity to try to estimate the SFR for galaxies lying in the redshift range $z \sim 0.4-1.5$. Photoionization models show that these are by no means ideal as star formation indicators, because their strength not only depends on the attenuation, which tends to be large in the UV, but also on the metallicity of the ionized gas. These may lead to large discrepancies between SFR estimated from \Ha\ and from the \OII\ lines \cite{Charlot02}. However, Kewley and her collaborators \cite{Kewley03} have developed techniques to allow corrections for both metallicity and reddening which bring the results of these two emission line techniques into much better agreement (see also the paper by Kewley, elsewhere in these proceedings).

\section{3. Physical Elements of SED Modelling}
A successful model for the SED of a starburst galaxy must bring together in one package a wide range of physics including the structure evolution and atmospheric models of stars, the full physics of ionized plasmas, a realistic dust model and a certain amount of molecular physics. In addition, the geometry of the dust with respect to the stars is fundamental in determining the shape and peak of the far-IR re-emission, as discussed by Popescu \& Tuffs (this volume). Ideally we should seek to develop an understanding of the geometry from the constraints offered by the other physics of the problem. Finally, the radiative transfer problem must be solved for the complex multi-phase nature of the interstellar medium. Few, if any, models have all these physical elements included at the same time.

\paragraph{\bf Stellar Spectral Energy Synthesis} There are a number of publicly available codes which give consistent results on the SED of the stellar component, for whatever choices of the IMF are preferred. These include the Starburst~99 code of Leitherer \cite{Leitherer99}, the P\'egase code \cite{Fioc97}, the Bruzual and Charlot code \cite{Bruzual03} and the code by  Kodama \& Arimoto \cite{Kodama97}. These codes differ mainly on their predictions in the EUV, since here they rely on theoretical atmospheric and mass-loss models in which a good deal of progress has been made recently. In this respect the Starburst~99 code is probably the state of art \cite{Smith02}. The major strength of the Bruzual and Charlot code is in the synthesis of the older stellar populations.

\paragraph{\bf Ionized Gas Physics} For the ionized plasma, two major codes have been used in SED modelling, the Cloudy code\cite{Ferland96}, and the Mappings III code \cite{Sutherland93, Dopita00a, Dopita00b}. Both of these include all relevant gas-phase physics and a complete model for the absorption and re-emission of radiation by dust grains in the nebula. These codes provide both the emission line SFR diagnostics discussed above, as well as giving estimates of both the chemical abundance and the ionization parameter in the ionized gas\cite{Dopita00a, Kewley02b}. The ionization parameter provides us with an independent physical constraint on the geometry of the gas and dust with respect to the central stars.

\paragraph{\bf Grain Composition and Size Distribution} Most models use a dust model consisting of both a carbonaceous and silicaceous component. Apart from composition, the grain size distribution and the grain composition are the principal factors which determine the  wavelength-dependence of the absorption and scattering processes. Most models use a grain size distribution and a composition which yield results consistent with observations of the extinction curve, and which also attempts to match the constraints on the cosmic abundances of the atomic species locked up the the grains in local and Magellanic Cloud environments. Examples include the older D\'esert et al \cite{Desert90} the more recent Weingartner \& Draine model \cite{Weingartner01a, Weingartner01b} or the models of Dwek and his collaborators \cite{Dwek04}.

The far-IR re-emission process depends critically on the grain size distribution, since small grains are subject to large stochastic variations of temperature \cite{Li01, Draine01}. Most codes take this effect into account when computing the far-IR spectrum. The grain size distribution in the ISM results from the balance between the grain formation and destruction processes. It has usually \citep{MRN77} been represented by a power law over a wide range of sizes, $a$;
\begin{equation}
dN(a)/da = k a^{-\alpha} \quad a_\mathrm{min} \le a \le a_\mathrm{max}. \label{MRN}
\end{equation}
where $\alpha$, $a_\mathrm{min}$ and $ a_\mathrm{max}$ are derived by an empirical fit to the scattering and extinction in the local ISM. A more complex distribution has been suggested on physical grounds \citep{Weingartner01a}, and this takes into account both silicate and carbon-containing grains with different grain-size distributions. For the carbonaceous grains, the polycyclic aromatic hydrocarbon grains (PAHs) are treated as providing an additional component on a continuous  distribution of grain sizes. Even these distributions can be approximated by power-laws over a wide range of radii. 

Grain shattering has been shown to lead naturally to the formation of a power-law size distribution of grains with $\alpha \sim 3.3$ \citep{Jones96}, observationally indistinguishable from the MRN value, $\alpha \sim 3.5$. Such a grain size distribution can also hold only between certain limits in size,the smaller size being determined by natural destruction processes such as photodestruction, and the upper limit being determined by limits on the growth by condensation and sticking. To capture these elements of the physics, Dopita et al \cite{Dopita04} have adopted a modified grain shattering profile with the form;
\begin{equation}
dN(a)/da = k  a^{-\alpha}
\frac{e^{-(a/a_\mathrm{min})^{-3}}}{1+e^{(a/a_\mathrm{max})^3}}. \label{GSprof}
\end{equation}

Although it is certainly true that the size distribution is a function of the environment, it is very dangerous to make the grain size distribution a free parameter, since this could enable one to fit the far-IR emission peak with entirely the wrong geometry of the gas with respect to the stars. It is better to take the approach pioneered by Dwek, and use the constraints offered by the UV attenuation by the dust simultaneously with fitting the far-IR peak. However, in such an exercise we must also take account the turbulent fractal structure of the ISM dust screen, which provides a greater than expected opacity in the IR, but which gives a smaller than expected attenuation in the UV, \emph{see} Fischera \cite{Fischera03, Fischera04, Fischera04b}.

\paragraph{\bf Grain Destruction} Since, in this conference, the possibility that environmental effects change the nature of the grains, the grain size distribution, and the relative abundance of the PAH molecules has been discussed at some length, let us now consider these issues. The life-cycle of dust grains in the interstellar medium must, in large measure, represent a balance between grain destruction and shattering in the fast shocks found in the hot and warm phases of the ISM, and the processes which build up grains in the dense molecular clouds. Dust evolution therefore depends upon the mass transport between these various phases. The presence or absence of the dust grains in a particular environment can be inferred by a measurement of elemental depletions in the interstellar medium.

Dwek \& Scalo \cite{Dwek80} and McKee \cite{McKee89}, have developed the evolutionary equations which describe these processes. These are simplified here. Let $M$ be the mass of a particular grain-forming element in the ISM, and suppose a fraction $\delta $ of this element is locked up into dust in the interstellar medium. In the hot phase, the evolution of $\delta $ will depend on the characteristic timescale needed to destroy the grains in the (warm) low-density phase through supernova explosions, $\tau _{\rm SNR}$, the timescale needed to regenerate the dust in the cold dense phase$\tau _{\rm form}$, and the time taken to transport the dust from the (cold) dense to the (hotter) less-dense phase, $ \tau _{\rm C\rightarrow H}$. Assuming that the growth time of dust grains in the cold phase is very short compared with the timescale of mass transport between the cold and warm phases (or its inverse), the dynamical evolution of the dust mass in the hot phase is:
\begin{equation}
{d(M_{\rm H} \delta_{\rm H}) \over {dt}} = {{M_{\rm C} \delta_{\rm C}} \over  {\tau _{\rm C\rightarrow H}}}   -  {{M_{\rm H} \delta_{\rm H}} \over {\tau _{\rm SNR}}}  
\end{equation}
and in equilibrium;
\begin{equation}
{{\delta_{\rm H}} \over {\delta_{\rm C}}} =  {{\tau _{\rm SNR} M_{\rm C}} \over {\tau _{\rm C\rightarrow H} M_{\rm H}}}. \label{Grain Eq.}
\end{equation}
In other words, the difference in the dust fractions between the two phases depends simply on the relative size of the reservoir of dust in the cold phase and ratio of the time taken for the dust to escape into the hot phase to the time taken to destroy it in the hot phase.

Now, what determines the timescale of destruction $\tau _{\rm SNR}$? This is the mean interval between the passage of fast shocks which result in sputtering of grains or in the destruction of PAH molecules. The sputtering rate is a rather complex function of temperature, and threshold energy\cite{Draine79} but can be rather well fitted by an equation of the form \cite{ Dwek96}: 
\begin{equation}
\frac{da}{dt}=AnT_{6}^{-1/4}\exp \left[ -BT_{6}^{-1/2}\right]  \label{Sputter}
\end{equation}
where $T_{6}$ is the gas temperature in units of 10$^{6}$K and $A$ and $B$ are constants. For graphite, $A= 6\times 10^{-6}\mu m$~yr$^{-1}$ and $B=3.7$ while for silicate $A=1.8\times 10^{-5}\mu m$~yr$^{-1}$ and $B=4$. The sputtering lifetime for average size grains ($a\sim 0.1\mu m)$ is of order 10$^{5}$~years in fast shocks. This is comfortably shorter than the grain braking timescale, so that only the very largest grains can survive. Depending on the magnetic field configuration, grains start to be destroyed (rather than simply shattered) in shocks faster than 100\kms, and the process is complete by $\sim 400$\kms.

How often do shocks this fast pass through the warm phase of the interstellar medium? We can estimate this by asking at what size and shock velocity do supernova remnants become dominated by cooling, after which they rapidly slow down. The end of the adiabatic (Sedov-Taylor) phase is signalled by the cooling timescale in the shell becoming less than the dynamical expansion timescale; $\tau_{\rm cool} < \tau_{\rm dyn}$.The cooling timescale in a strong shock can be approximated \cite{DopSuth03} by:
\begin{equation}
\tau_{\rm cool} =200v_{100}^{13/3}/Zn {\rm ~yr}
\end{equation}
 where $v_{100}$ is the shock velocity, $v_{\rm s}$, in units of 100\kms, the chemical abundance $Z$ is with respect to solar, and the pre-shock density is $n$ cm$^{-3}$. The Sedov Taylor solution for the SNR expansion gives:
\begin{equation}
\tau_{\rm dyn}={\rm 3.9\times10^5}E_{51}^{1/3}n^{-1/3}v_{100}^{-5/3} {\rm years}  \label{Dyn}
\end{equation}
were $E_{51}$ is the kinetic energy of the explosion in units of $10^{51}$ ergs. Therefore, we can infer that the SNR becomes radiative when:
\begin{equation}
v_{\rm 100}={\rm 3.5}E_{51}^{1/6}n^{-1/9}Z^{1/6}.  \label{Rad}
\end{equation}
This is interestingly close to the velocity at which sputtering and grain destruction is complete. The radius at which this condition is met is found by noting that $t=2R/5v_{\rm s}$ and using this to eliminate  time in the Sedov equation for the radius:
\begin{equation}
v_{\rm 100}={\rm 8.0}E_{51}^{1/2}n^{-1/2}R_{\rm 10}^{-3/2} {\rm \kms}   \label{Sed}
\end{equation}
where $R_{10}$ is the radius in units of 10pc. Thus, from eqns. \ref{Rad} and \ref{Sed} we have that the SNR becomes radiative when:
\begin{equation}
R_{10}^3={\rm 5.2}E_{51}^{0.66}n^{-1.22}Z^{-0.33}.  \label{Radius}
\end{equation}
In a disk-like galaxy, the gas with surface density $\Sigma_{\rm g}$ $\msolar $pc$^{-2}$ is confined within twice the scale height, $H$. The surface rate of supernova events is proportional to the surface rate of star formation , $\Sigma_{\rm SNR} \propto \Sigma_{\rm SFR}$, and the observational Schmidt Law \cite{Kennicutt98,Misiriotis04} gives $\Sigma_{\rm SFR} \propto \Sigma_{\rm gas}^{1.4}$. These relationships appear to hold in the case of starburst galaxies as well. Thus the timescale for the gas to be processed through a supernova shock is:
\begin{equation}
\tau_{\rm SNR} \propto H/(\Sigma_{\rm SNR} R_{10}^3) \propto  E_{51}^{-0.66} \Sigma_{\rm gas}^{-0.18} H^{-0.22}Z^{0.33} \sim {\rm const.} \Sigma_{\rm gas}^{-0.4} Z^{0.33} \label{Tauproc}
\end{equation}
because, in a self-gravitating layer, the scale height is set by the velocity dispersion of the disk gas in the vertical plane, $v_z$, giving $H=v_z^2/\pi {\rm G} \Sigma_{\rm gas}$, and in disk galaxies the observed \HI\ vertical velocity dispersion is observed to be almost constant at between 7 and 10 \kms  
\cite{Kruit1982, Shostak1984,Kim1999,Sellwood1999}.  In disk galaxies,  as $\Sigma_{\rm gas}$ decreases with increasing radius,so does $Z$, and therefore the two terms in eqn. \ref{Tauproc} almost cancel each other, giving the final result that $\tau_{\rm SNR}  \sim  {\rm const.} $

The fact that the SN processing timescale is virtually constant is a remarkable and perhaps even a counter-intuitive result. It may be giving us a profound clue about the process of self-regulation of star formation in galaxies. Since the supernovae act to turbulently stir the ISM, and since the ISM density structure is thought to originate from the turbulent cascade of energy from large scales to small scales, and since this turbulent cascade is thought to set up the conditions for new star formation, then the Schmidt law itself may simply be the expression of this turbulent cascade process with constant $\tau_{\rm SNR}$.

Return now to the effect that this has on the grain population. It is clear from equation \ref{Grain Eq.} that, if the SNR rate is also determining the timescale of large-scale mixing, and therefore the mass-exchange rate between the various phases, the ratio of the two timescales may not change much with environment. Therefore if we are to deplete the dust fraction in the hotter phase, we must rely on decreasing the mass fraction of the cold phase relative to the hotter phase as a function of environment.         

\paragraph{\bf PAH Physics} The PAH molecules having a molecular weight similar to coronene (${\rm C}_{24}{\rm H}_{12}$), are very important in absorbing UV photons, and fluorescing in the IR in a number of strong and broad emission line features in the wavelengh range between 3.29\mum\ and 20\mum. Being carbonaceous, and somewhat graphitic in their molecular structure, the PAH molecules have strong absorption in the 2175 \AA\  band. The absence of this absorption feature in the attenuation curve of the light of starburst galaxies  \cite{Calzetti01}. as well as the relative weakness of the IR PAH features in some starburst galaxies such as Arp220 may be evidence of the photodestruction of PAH molecules in environments with a strong UV or EUV field.      

This is supported by observations in our own galaxy. Where detailed measurements have been made of the distribution of PAH emission within compact \HII\ regions, it is found that the PAH avoids the ionized regions, and is instead found in a narrow zone of emission (the photo-dissociation region) beyond the outer boundary of the ionized gas \cite{Burton00}. This is clear evidence that PAH molecules cannot survive for significant lengths of time in the hostile environment in the ionized zones of an \HII\ region. This is to be expected, because the photodissociation timescales in this region are very short. However, PAHs can survive, and are excited into emission in the photodissociation regions (PDRs) outside the ionized \HII\ region.

The physics of the photodestruction process is rather uncertain, and so this leads to uncertainties in the form of the SED where the PAH features are strong. The most likely destruction process is by PAH charging. The $\pi$ electrons associated with the benzene rings are relatively easy to remove. Removal of the first causes buckling of the C-C skeleton with strong 6.2\mum\ and $7.7-7.9$ \mum skeletal deformation bands and both a broadening and change in relative intensities of the other IR emission features. Indeed, charged PAHs are an essential feature of any model which can reproduce the observed PAH IR emission spectrum. Most models use the ``astronomical'' PAH absorption properties (based largely on the measured properties of coronene) by Li and Draine \cite{Li01}. PAH photodestruction appears to be caused by removal of two or more electrons, which requires photons harder than about 14eV. Thus, the simplest PAH destruction model model is that the presence of  hard photons leads to rapid photodissociation in the ionized gas. In this case, the PAHs are destroyed in the ionized gas and their C is returned to the gas phase. Evidence for this can be adduced from the fact that, in \HII\ regions \cite{Garnett99}, the measured gas-phase log[C/O] has its solar value at solar log[O/H], using the new solar abundances \cite{Asplund00a, Allende01, Allende02}. This argues for the complete photodestruction of C-containing grains in \HII\ regions.

A somewhat more sophisticated approach might be to say that PAHs are destroyed whenever the absorption weighted mean radiation field (not necessarily ionizing) results in a photodissociation timescale which  is short compared to the residence time of the PAHs in that radiation field. In this case we would compare photodissociation timescales with dynamical timescales to obtain a criterion for survival. 

A third approach posits the possibility that photodissociation (by the ejection of an acetyleneic group) is countered by repair through accretion of carbon atoms \cite{Allain96a,Allain96b}. However, such a repair mechanism could only occur in photodissociation regions, not in the ionized gas where such groups would be rapidly photo-dissociated.                                                 

\paragraph{\bf An ISM/Dust Model for Starburst Galaxies} We \cite{Dopita04} have used a grain shattering profile of the form of eqn. (\ref{GSprof}) with both graphitic and silicaceous components. The grain size limits for both graphitic and silicaceous grains are set to $a_\mathrm{min}=0.004$\mum and $a_\mathrm{max}=0.25$\mum. The somewhat smaller size cutoff than the MRN distribution provides
a stronger dust re-emission continuum below 25 \mum, which is required to fit the observed IRAS colors of  starbursts \cite{Rush93}.  The constant k is determined by the total dust-to-gas mass ratio, which is determined by the depletion of  the heavy elements onto dust. Our models use a solar abundance set as modified by the latest abundance determinations \cite{Allende01,Allende02,Asplund00a,Asplund00b}. The depletion factors are those used by  \citet{Dopita00a} for starburst and active galaxy photoionization modeling  and are similar to those found by \citet{Jenkins87}  and  \citet{SavageS96} in the local ISM using the UV absorption lines to probe various local lines of sight. 

When present, the PAHs are set at an abundance which uses 110ppm of Carbon. Consequently, the carbonaceous  grain component is present at only rather low abundance; equivalent to 69ppm of Carbon. The PAHs are assumed to be destroyed whenever there are appreciable numbers of photons with energies greater than 14.5eV present, \emph{i.e.} effectively throughout the ionized portions of the \HII\ regions. This is all consistent with the observed C/O and O/H measurements of gas-phase abundances in \HII\ regions in the UV made by Garnett et al. \cite{Garnett99} for which both the C/O and O/H ratios agree with their solar values on the new solar abundance scale. This indicates that the depletion factor of C in the ionized gas is similar to that of O, which is known to be small.

\begin{figure}
{\includegraphics[height=.35\textheight]{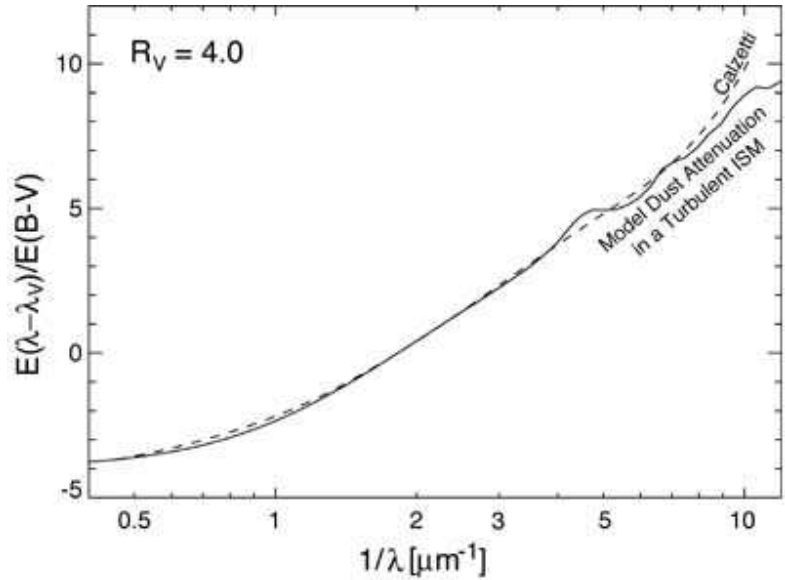}}
  \caption{The interstellar dust attenuation curve computed for our dust model  and for a turbulent ISM with a log-normal density distribution (Dopita et al. 2004 \cite{Dopita04}) (solid line) is compared with the empirical Calzetti attenuation curve derived for starburst galaxies. The two are identical within the errors.} \label{Calzetti}
\end{figure}

The destruction of PAHs in the ionized and diffuse phases of the ISM will largely remove the 2200\AA\ bump from the UV attenuation curve. In order to compute this attenuation we have to take account the turbulent fractal structure of the ISM dust screen \emph{see} Fischera \cite{Fischera03, Fischera04, Fischera04b}. The cloudy fractal nature of the foreground screen flattens the attenuation curve relative to the extinction curve (which we cannot in any case observe in starburst galaxies, since we cannot make measurements of the extinction in lines of sight to individual stars). When this is done, our grain model provides an excellent fit to the Calzetti \cite{Calzetti01} attenuation curve (\emph{see} Fig\ref{Calzetti}). This curve has been derived empirically from an ensemble of observations of starburst galaxies, and which has been widely used in the literature.

\paragraph{\bf Geometrical Concerns} The dust grain temperature distribution, and therefore the shape and peak of the far-IR peak depends critically on the geometrical relationship between the dust grains and the stellar heating sources assumed in the model. Models with warmer far-IR colors will have a more compact disposition of gas with respect to the stars. The difficulty here is that, in any starburst model, these geometrical relationships are not determined \emph{a priori}. 

In the semi-empirical modelling of Dale and his collaborators (\cite{Dale01, Dale02}, and this conference), the SED of disk and starburst galaxies were suggested to form a one-parameter family in dust temperature. This suggested that starburst galaxies have hotter dust temperatures. Lagache et al \cite{Lagache03} (again empirically) have suggested that the luminosity controls the form of the SED. Both of these may be true to some extent, since IR luminous galaxies have greater rates of star formation.

The French group \cite{Galliano03} take the simplest approach of approximating the starburst by a spherical \HII\ region and clumpy dust shell around the central star forming region. A more sophisticated approach is taken in the GRAZIL code by the Padova/Trieste group \cite{Silva98, Granato00, Panuzzo03}. Their starburst model uses a spherical geometry with King profiles, and they allow for the formation of clusters of stars in molecular complexes, and their subsequent escape from these regions.

In the rather sophisticated models of Takagi et al \cite{Takagi03a, Takagi03b}, a mass-radius relationship for the star formation region of $r_i/{\rm kpc} = \Theta (M/{10^9 {\rm \msolar}})^{1/2}$ is adopted along with a stellar density distribution given by a generalised King profile. The parameter $ \Theta $ is a compactness parameter which expresses the degree of matter concentration, and is related to the optical depth of the dust through which the starburst region is seen. For a sample of ultra-luminous starbursts, they find that, while most conform to a constant surface brightness of order $10^{12}\lsolar$kpc$^{-2}$, there are a few objects with surface brightnesses roughly ten times larger than this, which they ascribe to post-merger systems.

In a starburst, the individual stars form in clusters at the cores of dense molecular complexes, and initially the geometry of the gas with respect to the dust is determined by the evolution of the mass loss bubble / \HII\ region around individual clusters. It is only at later times that these bubbles may link up to create a large-scale \HII\ region complex driven by collective effects. In dense starbust regions, the shape of the SED is therefore determined more on the scale of individual \HII\ regions, rather than on a galaxian scale. The size of each \HII\ region is therefore a function of both the mass of the central cluster and its age. To provide the correct geometry, we therefore need to understand the time evolution of individual \HII\ regions. An attempt to do this was recently described by Dopita et al. \cite{Dopita04}, and this work forms the basis of the following section.

\section{4. HII Region Evolution in Starburst Galaxies}
\begin{figure}
{\includegraphics[height=.45\textheight]{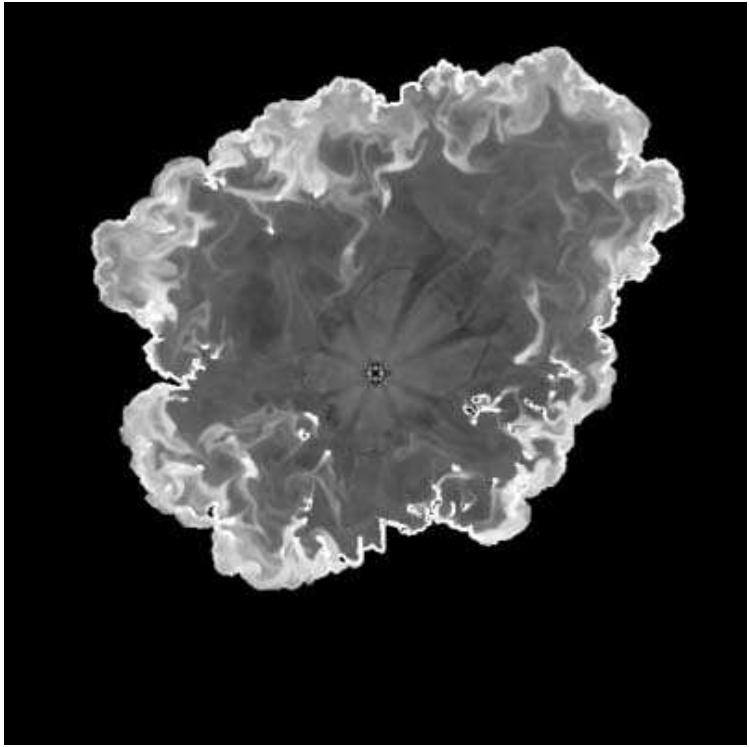}}
  \caption{The radiative structure of a 2-D radiative mass-loss blown bubble evolving in a turbulent ISM with a log-normal density distribution having a mean atom density of unity. This has been computed using the PPMLR code developed by  \cite{Sutherland02}. It shows a number of features seen in real \HII\ regions such as elephant-trunk clouds, and demonstrates how different is the evolution compared to the case of a uniform density medium. In a log-normal density distribution, the \HII\ mass-loss bubble expands more rapidly than in a uniform medium, and it leaves a number of high emission-measure regions having relatively high dust temperatures close to the exciting star(s).} \label{Fig2}
\end{figure}

In what follows, one should be aware that the problem of the evolution of the size of \HII\ regions in the interstellar medium is by no means trivial. Although we will assume that the ISM is uniform, and of a uniform pressure, such is certainly not the case in real objects. The turbulent multi-phase structure of the ISM profoundly modifies the evolution of \HII\ regions. An attempt to model this, by evolving a source of uniform mass-loss in an isotropic log-normal density distribution (which mathematically approximates to the structures produced in a turbulent medium) is shown in \ref{Fig2}. Such models will be required in future modelling. For the time being, let us consider on the case of uniform pressure and density in the ISM.

The work of  Dopita et al. \cite{Dopita04} drew on the theory of the size distribution of \HII\ regions by Oey \& Clark  \cite{Oey97, Oey98}. In this, the \HII\ regions expand and evolve as stellar mass-loss and supernova driven bubbles for as long as their internal pressure exceeds the ambient pressure in the ISM. When this condition is no longer met, the \HII\ region is assumed to ``stall''. Of course, on reaching the stall condition, the \HII\ region  does not abruptly stop expanding, but a momentum-conserving expansion continues until interstellar turbulence  destroys the integrity of swept-up shell. The time taken to reach the stall radius turns out to be proportional to the stall radius, and the mass of the OB star cluster is important in determining the stall radius; \HII\ regions with low mass clusters stall early and at small radius for a given pressure or density in the ISM.

The radius $R$ and internal pressure,  $P$, of a mass-loss pressurised \HII\ region are given by  \cite{Oey97, Oey98};
\begin{equation}
R = \left(250 \over 308\pi\right)^{1/5}L_{\rm mech}^{1/5}\rho_0^{-1/5}t^{3/5} \label{R}
\end{equation}
and
\begin{equation}
P = {7 \over (3850\pi)^{2/5}}L_{\rm mech}^{2/5}\rho_0^{3/5}t^{-4/5}. \label{P}
\end{equation}
Here, $L_{\rm mech}$ is the mechanical luminosity of the mass-loss from the central stars, $\rho_0$ is the density  of the ambient medium and $t$ is the time. The particle density is given in terms of the ambient density by $n=\rho_0/\mu m_H$, and the ambient pressure $P_0=nkT_0$. Eliminating $t$ between equations \ref{R} and \ref{P}, and setting $P =P_0$ we obtain the following conditions for stall:
\begin{eqnarray}
P = {7 \over (3850\pi)^{2/5}}\left(250 \over 308\pi\right)^{4/15}{\left(L_{\rm mech} \over \mu m_Hn\right)}^{2/3}
{\mu m_Hn \over R^{4/3}} \label{Pstall1} \\ 
=nkT_0 = n_{\rm HII}kT_{\rm HII}. \label{Pstall2}
\end{eqnarray}

\paragraph {\bf Ionization Parameters} Equations \ref{Pstall1} and \ref{Pstall2} imply that, for any given pressure in the ISM, all stalled \HII\ regions have a common ratio of $L_{\rm mech}/n_{\rm HII}R^2$. To the extent that $N_* \propto L_{\rm mech}$, where $N_*$ is the flux of ionizing photons from the central stars, then all stalled \HII\ regions will be characterized by a common value of the ionization parameter, ${\cal U} =N_*/4\pi c R_{\rm in}^2 n_{\rm HII}$. This ionization parameter is very important in the theory of \HII\ region spectra since it is one of the three parameters which determine the emission line spectrum of the \HII\ region. The other two are the input spectrum of the exciting star(s) and the gas-phase metallicity.

In fact, even for \HII\ regions which are not stalled, there is a unique value of the ionization parameter in the \HII\ region. This is because the pressure in the stellar winds at the outer boundary of the mass-loss bubble determines the equilibrium pressure in the \HII\ region plasma. Thus, $P_{\rm w} = L_{\rm mech}/4 \pi R_{\rm w}^2v_{\rm w} =P_{\rm HII} = n_{\rm HII}kT_{\rm HII}$, where $v_{\rm w}$ is the mean stellar wind velocity measured at the radius of termination of the free-wind region $R_{\rm w}$. This free-wind boundary is visible in Fig \ref{Fig2} as the boundary of the inner region where the cross-shaped cooling artifacts in the free-wind resulting from the finite wind injection region give way to the shocked hot turbulent gas filling the rest of the bubble. In general we can take $R_{\rm w}=\alpha R$ with $\alpha \sim 0.3$. Thus:
\begin{equation}
{\cal U} =\alpha^2kT_{\rm HII}N_*v_{\rm w}/ cL_{\rm mech} \label{U_HII}
\end{equation}
Since the \HII\ region temperature is always $T_{\rm HII} \sim 10^4$K, then, for any central star cluster at a particular time in its evolution, the ionization parameter is determined solely by the instantaneous ratio of $N_*v_{\rm w}/ L_{\rm mech}$. Thus, for coaeval clusters which are massive enough so that stochastic variations in these quantities are unimportant, the ionization parameter will be essentially independent of cluster mass and is only a function of cluster age. All the quantities needed to compute the ionization parameter, are given in the STARBURST99 output files.

\begin{figure}
  {\includegraphics[height=.4\textheight]{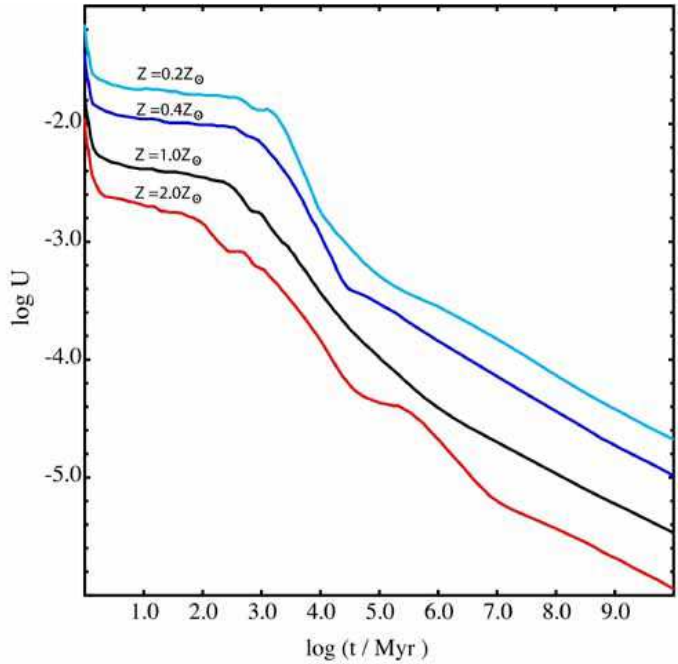}}
  \caption{The time dependence of the ionization parameter of the \HII\ region as a function of metallicity computed for a cluster of mass $3\times10^3\msolar$ from STARBURST99, with a mechanical energy efficiency factor of 0.1. These computations include the evaluation of the self-consistent size evolution of the \HII\ region.} \label{Fig3}
\end{figure}

Since the extended atmosphere of the exciting stars absorb a greater and greater fraction of the ionising photons as the metallicity increases, and at greater metallicity a greater fraction of the momentum flux in the radiation field is converted to mechanical energy in the winds, the ratio of $N_*v_{\rm w}/ L_{\rm mech}$ strongly decreases as the metallicity increases. The effect of this on ${\cal U}$  is shown in Fig \ref{Fig3}.

\paragraph{\bf Dust Temperatures} The determination of the dust temperature, or more properly the shape and the peak of the far-IR bump in starburst SEDs is not so straightforward as determining the ionization parameter. However, the general physics that drives the distribution of dust temperatures is readily understood.

We saw above how \HII\ region in high pressure (or density) environments will stall at a smaller radius, according to the Oey \& Clark theory \cite{Oey97, Oey98}. This means that the dust within the ionized gas of the \HII\ regions will be hotter when the ISM pressure is high. At the outer boundary of the \HII\ region we normally expect to find molecular clouds in approximate pressure equilibrium with the ionized gas. Therefore, the radiation field with energies below the Lyman Limit is absorbed by the gas and the dust in a rather thin photodissociation region (PDR) behind the ionization front which bounds these molecular clouds. At any point in the PDR, the local radiation field is determined only by the radiation field incident on the molecular cloud and by the optical depth up to that point. For a given grain size distribution and composition, the grain temperature distribution, and hence the far-IR re-emission function, is determined by the local radiation field. The global far-IR spectrum is the integral of the local far-IR spectra through the PDR. Hence the far-IR spectrum is determined only by the radiation field incident on the molecular cloud.

\begin{figure}
{\includegraphics[height=.5\textheight]{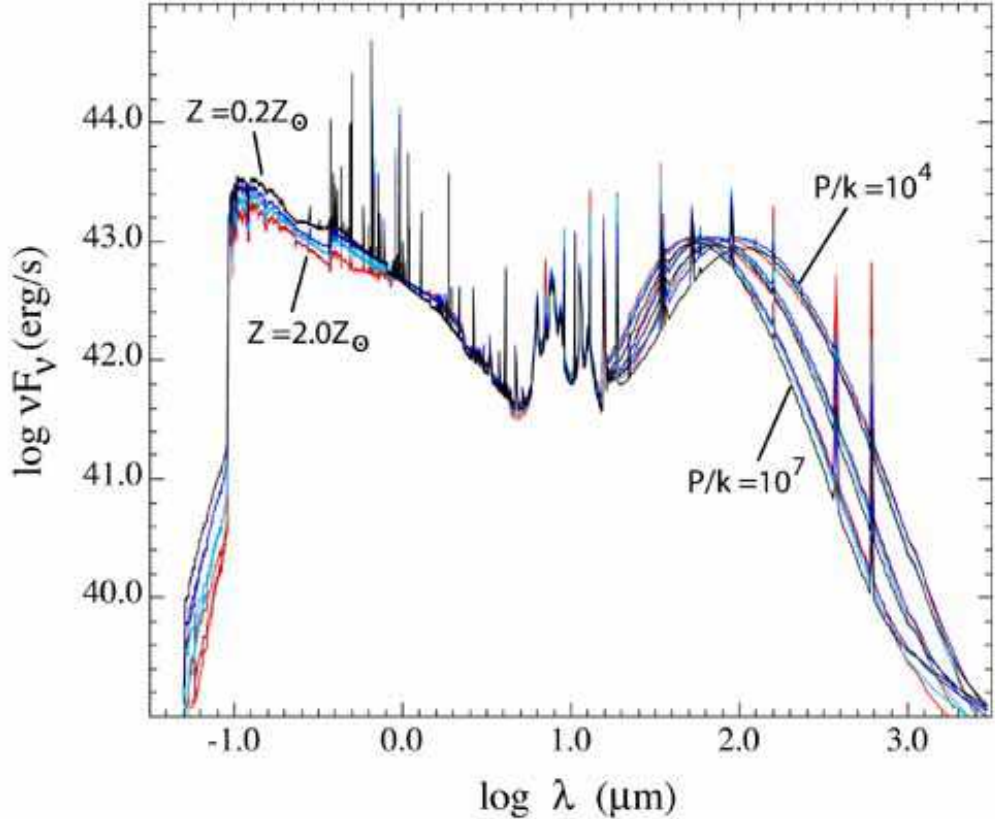}}
  \caption{Integrated starburst SEDs computed for metallicities 2.0, 1.0, 0.4 and 0.2 $Z_{\odot}$, for a fixed size distribution of \HII\ region diameters, and for a range of ISM pressures $P/k = 10^4, 10^6 \& 10^7$cm$^{-3}$K. Here the $<10$ Myr stellar population is assumed to reside within the \HII\ regions, and a $10-100$Myr population provides an unobscured foreground contribution. The molecular cloud shell around the \HII\ regions is assumed to be destroyed on an e-folding timescale of 8 Myr. Note that the optical and UV line and continuum spectra are strongly dependent on metallicity, but that the far-IR peak is a function only of the pressure in the ISM. Note also the invariance of the PAH emission features in the mid-IR. } \label{Fig4}
\end{figure}

The result of this is that smaller \HII\ regions, which are preferentially encountered in region of the ISM with higher pressure will be characterised by hotter far-IR dust emission spectra. Provided that the PDR remains thin in comparison with the radius of the \HII\ region, the far-IR spectrum of the ensemble of \HII\ regions remains a function only of the pressure. The abundance of the heavy elements can only make a difference when the changing abundance drives a changing grain size distribution, when the changing ratio of photon energy to mechanical energy produced by the central stars leads to a changing size distribution in the \HII\ region population, or when the opacity of the grains falls so low that the spherical divergence of the radiation field through the PDR becomes important. This dependence of the far-IR spectrum on pressure is illustrated in Fig \ref{Fig4}.

The invariance of the PAH features in these models is interesting. This occurs despite the fact that the abundance of changes by a factor of nearly 30, and the relative abundance of PAH molecules relative to silicaceous grains decreases strongly at low metallicity. The reason for this is that at the wavelength where they strongly absorb, the PAH molecules dominate the total dust and molecular opacity. They therefore convert a fixed fraction of the bolometric luminosity of the star to PAH emission features. In order to decrease the relative strength of the PAH features at low metallicity, we need to ensure that the molecular clouds become ``leaky'' to the incident radiation, which requires that the column density in the molecular clouds does not exceed $\sim 3 \times 10^{21}$ H atoms cm$^{-2}$.

\section{5. Pan-spectral fits to Starburst Galaxies}\label{SEDfits}
\paragraph{\bf SED Fits to ULIRG Galaxies} In order to test these ``multiple \HII\ region'' models for the SEDs of Starburst Galaxies, Reuland et al. \cite{Reuland05} have collected data from the literature for 41 ultra-luminous infrared galaxies (ULIRGs) in the local universe. The data consist of UV/optical fluxes from the {\it Third Reference Catalogue of Bright Galaxies} v3.9, aperture-corrected JHK-band photometry from Spignolio \cite{Spignolio95} and 3-1500 \mum data from Klaas et al.\cite{Klaas97, Klaas01} and from Spoon \cite{Spoon04}. The cosmological model used to derive luminosity distance is the concordance cosmology  with $\Omega_M=0.27, \Omega_{\Lambda}=0.73$ and $H_o=71$km s$^{-1}$ Mpc$^{-1}$ \cite{Spergel03, Tonry03}.

\begin{figure}
{\includegraphics[height=.5\textheight]{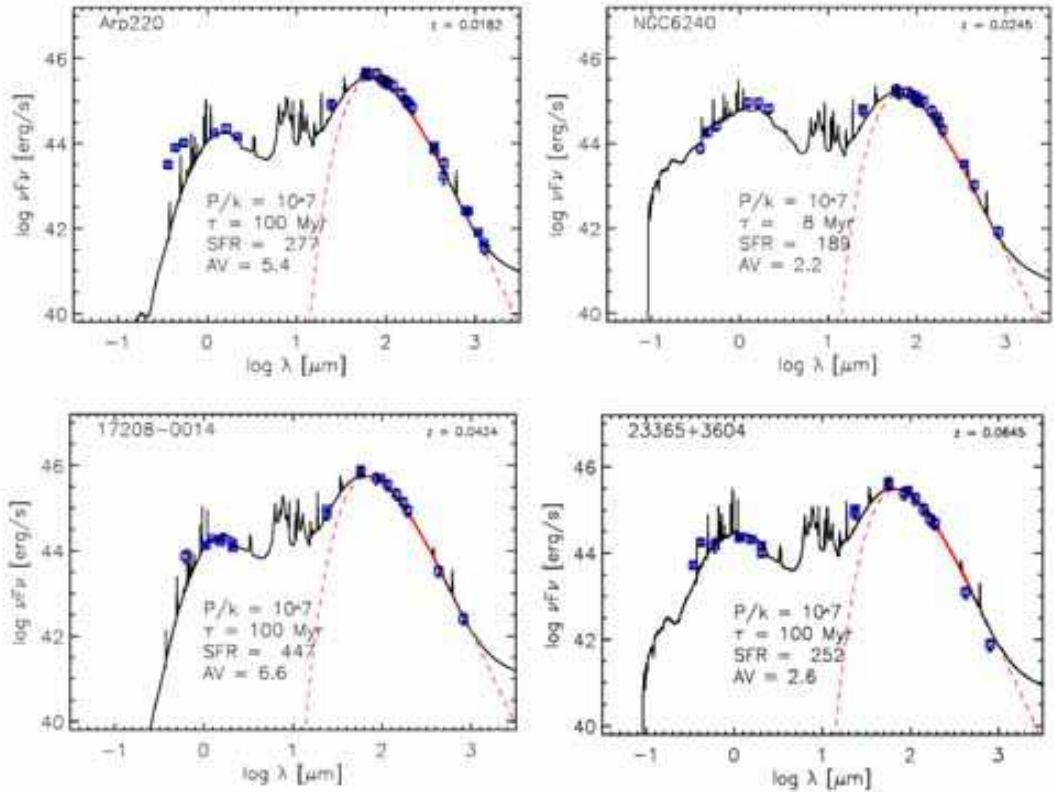}}
  \caption{Here we show the fits to the SEDs of a few well-known ULIRGS in the local universe, scaled to a star formation rate of $1.0 \msolar $yr$^{-1}$. The shape of the far-IR bump is used to fit the pressure; $P/k$. The visible and UV part of the spectrum is fitted by a combination of foreground screen attenuation with an effective $A_{\rm v}$, and a molecular cloud dissipation timescale, $\tau$Myr. Finally, the overall scaling factor provides the estimate of the total star formation rate. }  \label{Fig5}
\end{figure}

The variables characterising the fit are:
\begin{itemize}
\item{The ISM pressure, $P/k$~cm$^{-3}$K. As described above, this determines the temperature distribution of the grains, and hence the position and shape of the far-IR dust re-emission peak. It serves the same role as the compactness parameter of Takagi et al \cite{Takagi03a, Takagi03b}}
\item{The attenuation of the foreground dust screen expressed in equivalent optical magnitudes of extinction, $A_{\rm v}$. The extinction computation is not fully self-consistent, because we do not compute the far-IR emission arising from this absorption. However, this will be unimportant provided that the covering factor of the molecular clouds around the star formation region is large. This should be a good approximation because, as discussed above, the dust obscuration is observed to increase both with galaxian mass \cite{Kewley02}, and with the absolute rate star formation \cite{Buat99, Adelberger00,Dopita02, Vijh03}.}
\item{The molecular cloud destruction timescale, or, equivalently, the timescale taken for stars to escape from the dense region of star formation and molecular clouds, $\tau$Myr. This effectively determines the fraction of UV / visible radiation which is intercepted by dust in dense clouds, and which is therefore re-emitted in the far-IR. This number effectively determined the intrinsic ration of UV to far-IR, while the previous factor determines the shape of the UV / visible SED.}
\item{ The star formation rate (SFR, $ \msolar $yr$^{-1}$). The model spectra are scaled to an absolute star formation rate of  $1.0 \msolar $yr$^{-1}$, so that the SFR is simply the inverse of the factor by which the observed SED has to be scaled to fit the theoretical models.}
\end{itemize}

The result of this fitting exercise is shown in Fig \ref{Fig5} for four well-known ULIRGS. The remaining 37 objects give fits of comparable quality. The largest deviations are seen in the far-UV, but this is most likely due to the presence of a small number of relatively unobscured stars. These make a negligible contribution to the Bolometric flux, and hence to the estimated total SFR.

\paragraph{\bf Implications for hi-z Starbursts} All of the fits ULIRGS are characterized by high pressures, $logP/k > 6$~cm$^{-3}$K, and therefore have far-IR bumps which peak below 100\mum . This result is not unexpected, because the star formation region can only be as compact as it is in ULIRGS if the star formation is occuring in a very high pressure and high density environment. The pressure can be indirectly inferred from measurements of pressure in the \HII\ regions though measurements of the density-sensitive [S II] $\lambda 6717, 6731$ \AA\  doublet in warm IRAS galaxies. This has been done by Kewley \cite{Kewley01, Kewley01b} and indeed, pressures $logP/k > 6$~cm$^{-3}$K are inferred for most objects. 

In the local universe, the ``dust temperature'' inferred from the modified Black-Body fits to the long wavelength side of the far-IR peak in starburst galaxies is observed to correlate with the absolute luminosity, which for these galaxies can be interpreted as proportional to the rate of star formation \cite{Blain04}. However, the same exercise applied to high-redshift submillimeter selected galaxies (SMGs) provides a similar correlation, but shifted to higher luminosity. At a given luminosity, the dust in SMGs is about 20K cooler than in ULIRGs in the local universe, and at a given dust temperature, the SMGs are typically 30 times as luminous as their ULIRG counterparts.  

What does this mean? As reported above, Takagi et al \cite{Takagi03a, Takagi03b} had found that most ULIRGS have a constant surface brightness of order $10^{12}\lsolar$kpc$^{-2}$. Our results show that this corresponds to an ISM with a pressure of order  $logP/k \sim 7$~cm$^{-3}$K . These parameters probably characterise ``maximal'' star formation, above which gas is blown out into the halo of the galaxy and star formation quenched. Only mergers, which provide an additional ram-pressure confinement of the star formation activity may exceed this surface brightness. Thus, in order to scale the star formation up to the rates inferred for SMGs ($ \sim 1000-5000 \msolar $yr$^{-1}$), we must involve a greater area of the galaxy in star formation, rather than trying to cram more star formation into the same volume. For a typical value of $10^{13}\lsolar$kpc$^{-2}$, we require ``maximal'' star formation over an area of $\sim 10$kpc$^{2}$, and the most luminous SMGs require star formation to be extended over an area of at least $\sim 100$kpc$^{2}$. We can therefore conclude:
\begin{itemize}
\item{ SMGs are truly starbursts extended on a galaxy-wide scale, rather than the more confined or nuclear starbursts which characterise ULIRGs in the local or moderate-redshift universe.}
\item{Because of the greater physical extent inferred for the starburst region in the SMGs, the modelling parameters we have derived for local ULIRGS: pressure, molecular gas dissipation timescales, and line of sight attenuations  can probably be directly applied to modelling SMGs in the distant universe. This is good news indeed!}
\end{itemize}

\section{6. The Way Forward}
The modelling described here can be improved in a number of ways, before fully reliable SEDs of starburst galaxies can be obtained. In rough order of urgency:
\begin{itemize}
\item{ To perform the full radiative transfer and scattering calculations for the turbulent foreground screen of diffuse matter in order to correctly model the UV and the ratio of the UV to far-IR flux.}
\item{To correctly compute the SEDs of ultra-compact \HII\ regions around isolated OB stars, since these are likely to form a greater fraction of the \HII\ region population in high-pressure regions of the ISM.}
\item{To understand the process of the escape of stars from cluster environments into the general ISM.}
item{ To properly understand the life-cycle of PAHs in the ISM through both advances in theory, and through modelling the SEDs of individual star formation regions in nearby galaxies observed with Spitzer.}
\item{To quantify the X-ray emission mechanisms to allow star formation rates derived from X-rays to be properly calibrated.}
\end{itemize}


\begin{theacknowledgments}
 M. Dopita acknowledges the support of both the Australian National University and of the Australian Research Council (ARC) through his ARC Australian Federation Fellowship, and through the ARC Discovery project grant DP0208445. 
\end{theacknowledgments}


\bibliographystyle{aipproc}   


\end{document}